\documentclass[12pt,draftclsnofoot, onecolumn]{IEEEtran}
\ifCLASSINFOpdf
\else
\fi
\usepackage{enumitem}
\usepackage{algpseudocode}
\usepackage{mathtools}
\usepackage{algorithm}
\usepackage{setspace}
\usepackage{graphicx}
\usepackage{notoccite}
\usepackage{amssymb}
\usepackage{url}
\usepackage[table]{xcolor}
\usepackage{placeins}
\usepackage{amsfonts}
\usepackage{amsthm}
\usepackage{amsmath}
\usepackage{bbm}
\usepackage{subfigure}
\usepackage{multirow}
\usepackage{pdfrender}
\usepackage{color}

\begin{document}

\title{Dynamics of Quadrotor UAVs for Aerial Networks: An Energy Perspective}

\author{Wael Jaafar,~\IEEEmembership{Member,~IEEE,}
        Halim Yanikomeroglu,~\IEEEmembership{Fellow,~IEEE}
 
\thanks{Dr. Wael Jaafar and Prof. Halim Yanikomeroglu are with the Department of Systems and Computer Engineering, Carleton University, Ottawa,
ON, Canada, e-mails: \{waeljaafar\}\{halim\}@sce.carleton.ca.}
\thanks{This  work  is  supported  in  part  by  the  Natural  Sciences  and Engineering Research Council Canada (NSERC).}
\thanks{This work has been submitted to the IEEE for possible publication. Copyright may be transferred without notice, after which this version may no longer be accessible.}

}


\markboth{IEEE Wireless Communications Letters,~Vol.~XX, No.~XX, Month~Year}%
{Shell \MakeLowercase{\textit{et al.}}: Bare Demo of IEEEtran.cls for IEEE Journals}

\maketitle

\begin{abstract}
In this letter, we present a model for quadrotor unmanned aerial vehicles (UAVs), including control, communication, and wireless charging. In so doing, we investigate associated energy and battery dynamics. Indeed, energy and battery expressions are derived by leveraging motors' and battery electrical models. Through an experiment, their performances are evaluated for different parameters. 
The objective is to provide a simple yet practical model of quadrotor UAV consumed/harvested energy and battery dynamics for researchers conducting work on energy-efficient aerial networks.  
\end{abstract}



\IEEEpeerreviewmaketitle

\section{Introduction}

Unmanned aerial vehicles (UAVs) have been experiencing a boom in interest lately from industry and research. Indeed, several new applications that rely on UAVs have emerged in recent years in connection with the evolution of wireless networks into 5G and beyond. UAVs have been deployed for aerial security inspection, precision agriculture, traffic control, and package delivery. UAVs can also act as cellular base-stations (UAV-BSs) to provide connectivity to rural and disaster-hit areas. Hence, they are seen as a promising technology to profit businesses and help society. 

Despite all their promise, quadrotor UAVs have a major drawback: their flight duration is significantly limited and thus are unable to satisfy the requirements of all these emerging applications. This limitation is mainly due to existing lithium-ion polymer (LiPo) battery capacity. 
To bypass this limitation, several techniques have been proposed \cite{Galkin2018UAVsAM}. For instance, UAV swapping has been proposed, where an operating UAV with low power is repatriated to a dockstation and a freshly charged UAV substitutes it. The number of operating UAVs required depends greatly on the type of application, its duration and on the environment. Also, since the majority of UAVs nowadays are designed with external battery packs, the latter can be detached and replaced by charged batteries at an automated hotswapping dockstation. 
These solutions require the UAV to travel to a dockstation, which is time-consuming, especially for time-sensitive applications, such as rescue missions. An alternative solution that has been investigated to keep UAVs in the air for longer times is wireless power transfer (WPT). Two classes of WPT exist: electromagnetic field (EMF) charging and non-EMF charging. EMF-charging can transfer small amounts of energy on very short distances (a few centimeters), which cannot compensate for the ongoing power consumption of UAVs. Non-EMF charging by contrast uses photo-voltaic cells to harvest energy on longer distances. Unlike fixed-wing UAVs, small quadrotor UAVs cannot harness solar power. Instead, distributed laser charging (DLC) can be adopted \cite{Zhang2018}, where energy is harvested via a line-of-sight (LoS) link.   

In order to leverage UAVs for wireless applications, several challenges have to be addressed, including optimal UAV placement \cite{Hourani2014}, trajectory planning \cite{Zeng2016}, resource control \cite{Mei2018} and flight optimization \cite{Gong2018}. 
In considering these challenges, the issue of energy constraints is either absent, or only partially considered. In fact, in addition to energy requirements for communication related tasks, such as signal processing, radio-frequency (RF) circuits and amplification, energy requirements for propulsion in hovering or traveling to and from locations also need to be taken into consideration. Most research is limited to considering communication-related energy or motion energy from the perspective of flight duration or distance traveled, without considering the direct link between UAV energy and battery dynamics. Mathematical modelling of motion energy has received little attention. It is only recently that \cite{zeng2018energy} has proposed a propulsion power consumption expression for quadrotor UAV straight-and-level flight. However, a UAV travels in the 3D space with a complex trajectory, and changes its attitude according to the flight path and external forces, e.g. wind. For this reason, a simple yet accurate energy consumption model needs to be determined. To the best of our knowledge, no closed-form energy expression has been derived for UAV dynamics.          
Motivated by these discussions, and the apparent lack of a UAV model adapted to wireless applications, we present a simple quadrotor UAV model in this paper, where expressions of consumed/harvested energy are determined, and their relation to battery dynamics defined.


The main contributions of this letter are as follows. 1) From the aerial networks literature, we present a simple motion control, communication and WPT model for quadrotor UAVs. 
2) From the automation literature, the energy and battery dynamics expressions are obtained for quadrotor UAVs. They are illustrated afterwards through an experiment.  

The rest of this paper is organized as follows. In Section II, the quadrotor UAV model is presented. Section III details the energy expressions. In Section IV, the associated battery dynamics are given. Section V presents the experiment's results. Finally, Section VI concludes the paper.   

\section{Quadrotor UAV Model}
In this section, we present the motion control, communication and WPT models for the quadrotor UAV. 

\begin{figure}[t]
	\centering
	\includegraphics[width=200pt]{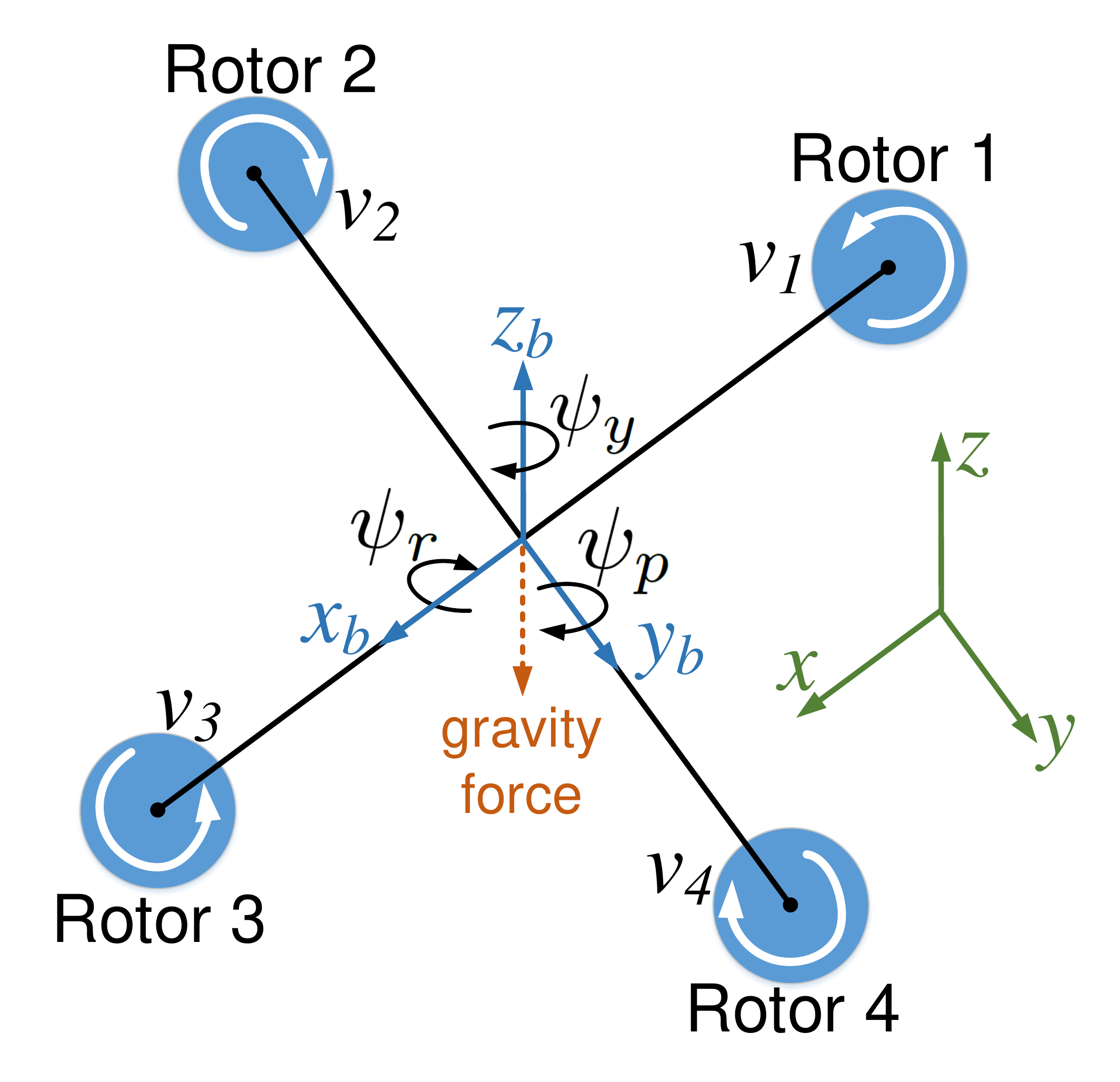}
	\caption{Quadrotor UAV.}
	\label{Fig:UAV}
\end{figure}

\subsection{Motion Control Model}
A quadrotor UAV is illustrated in Fig. \ref{Fig:UAV}. It has four rotors that control travelling and hovering actions. By adequately adjusting the velocities of rotors $v_{r}$ ($r\in \left\{1, 2, 3, 4\right\}$), the UAV can hover or travel horizontally/vertically. Let $(\psi_r,\psi_p,\psi_y)$ be the vector of roll, pitch and yaw angles capturing the attitude (i.e. orientation) of the UAV. These rotation angles are defined with respect to the body frame axis system ($x_b,y_b,z_b$).

\begin{figure}[t]
	\centering
	\includegraphics[width=200pt]{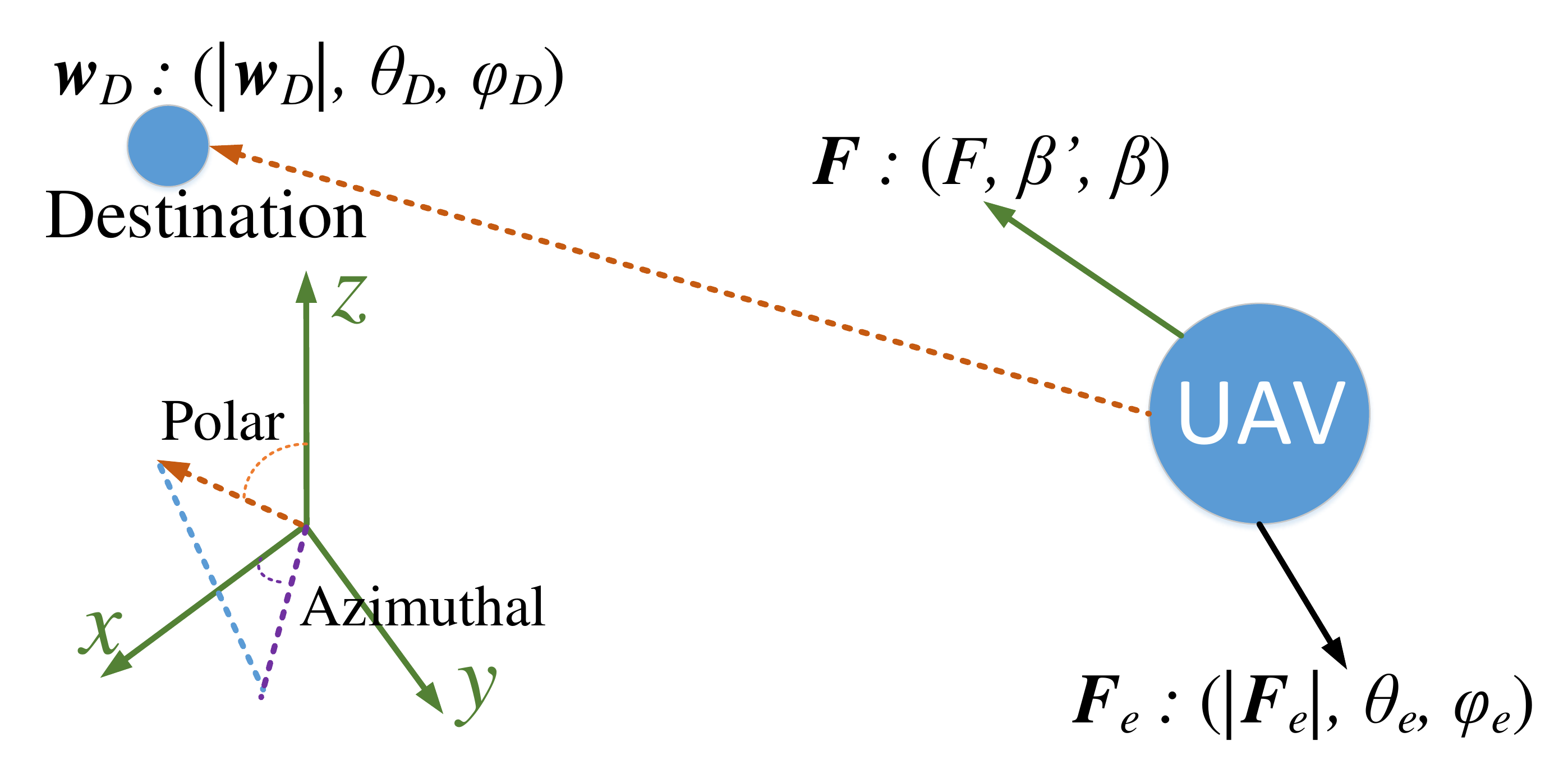}
	\caption{Movement towards a destination in presence of an external force.}
	\label{Fig:dest}
\end{figure}

According to Lemma 2 in \cite{Mozaffari2019}, when a UAV needs to fly towards a destination with coordinates $\textbf{w}_D=\left[x_D,y_D,z_D\right]^t$, in the presence of an external force $\textbf{F}_e=\left[F_{e,x},F_{e,y},F_{e,z}\right]^t$ (Fig. \ref{Fig:dest}), the flying attitude is given by
\begin{eqnarray}
\label{eq:attitude_D}
\psi_p^D&=&\text{cos}^{-1}\left( \frac{A\; \text{cos}(\theta_D) - |\textbf{F}_e|\; \text{cos}(\theta_e)}{F}\right),\\
\label{eq:attitude_D2}
\psi_r^D&=& \text{tan}^{-1}\left( \text{tan}(\beta) \cdot \text{sin}(\psi_p^D) \right),
\end{eqnarray} 
and $\psi_y^D=0$, where $F$ is the magnitude of the maximum force produced by the UAV, $|\textbf{F}_e|$ is the magnitude of $\textbf{F}_e$, $A=\left[ F^2 + |\textbf{F}_e|^2+ 2F |\textbf{F}_e| \text{cos}\left( \eta+\text{sin}^{-1} \Gamma \right) \right]^{1/2}$ with $\Gamma=\left( \frac{|\textbf{F}_e|}{F} \text{sin}(\eta) \right)$, $\theta_D=\text{cos}^{-1}\left(\frac{z_D}{|\textbf{w}_D|} \right)$, $\theta_e=\text{cos}^{-1}\left( \frac{F_{e,z}}{|\textbf{F}_e|}\right)$, $\eta=\text{cos}^{-1}\left(\frac{ \textbf{F}_e \cdot \textbf{w}_D}{|\textbf{F}_e| \cdot |\textbf{w}_D|}\right)$,
$\beta=\varphi_D- \text{sin}^{-1}\left( \frac{|\textbf{F}_e| \text{sin}{(\theta_e)} \text{sin}\left(\varphi_D-\varphi_e\right)}{F \text{sin}(\psi_p^D)}\right)$, $\beta'=\psi_p^D$, $\varphi_e=\text{tan}^{-1}\left( \frac{F_{e,y}}{{F}_{e,x}}\right)$, and $\varphi_D=\text{tan}^{-1}\left(\frac{y_D}{x_D} \right)$. All previously defined polar and azimuthal angles are based on the Cartesian-to-spherical coordinates transformation, with the center of the UAV as the origin of both coordinate systems.  

From Theorem 3 in \cite{Mozaffari2019}, travelling path from location $[0,0,0]$ to $\textbf{w}_D$ (assuming no obstacles along the path) can be broken down into six stages. In stages 1, 3 and 5, the UAV changes its attitude (i.e. roll and pitch angles). In stages 2 and 4, the UAV travels along a line with maximum and minimum acceleration respectively, while in stage 6, the UAV hovers at the destination. It is to be noted that while this six-stage approach yields a sub-optimal solution, it allows us to determine closed-form expressions for control inputs (i.e. rotor velocities), and hence simplifies its implementation in time-sensitive aerial networks. Accordingly, optimal rotor velocities to minimize control times are given by \cite{Mozaffari2019} 
\begin{eqnarray}
	\label{eq:sub1}
	v_{2}=0, \; v_{1}=v_{3}=\frac{1}{\sqrt{2}}v_{\rm{max}},\; v_{4}=v_{\rm{max}},\nonumber \\ 
	\text{if}\; t\in (0, \tau_1] \cup (\tau_5, \tau_6]\cup (\tau_{10}, \tau_{11}],  \\
	\label{eq:sub2}
	v_{4}=0, \; v_{1}=v_{3}=\frac{1}{\sqrt{2}}v_{\rm{max}},\; v_{2}=v_{\rm{max}},\nonumber \\  
	\text{if}\; t \in (\tau_1, \tau_2] \cup (\tau_6, \tau_7] \cup (\tau_{11}, \tau_{12}], \\
	\label{eq:sub3}
	v_{1}=0, \; v_{2}=v_{4}=\frac{1}{\sqrt{2}}v_{\rm{max}},\; v_{3}=v_{\rm{max}}, \nonumber \\ 
	\text{if}\; t \in (\tau_2, \tau_3] \cup (\tau_7, \tau_8]\cup (\tau_{12}, \tau_{13}],\\
	\label{eq:sub4}
	v_{3}=0, \; v_{2}=v_{4}=\frac{1}{\sqrt{2}}v_{\rm{max}},\; v_{1}=v_{\rm{max}}, \nonumber \\ 
	\text{if}\; t \in (\tau_3, \tau_4] \cup (\tau_8, \tau_9] \cup (\tau_{13}, \tau_{14}],\\
	\label{eq:sub5}
	v_{1}=v_{2}=v_{3}=v_{4}=v_{\rm{max}}, \; \text{if}\; t \in (\tau_4, \tau_5] \cup (\tau_9, \tau_{10}],\\
   \label{eq:sub6}
	v_{1}=v_{2}=v_{3}=v_{4}=\sqrt{\frac{|\mathbf{F}_e|}{4 \varrho}}, \;\text{if}\; t > \tau_{14},
\end{eqnarray}
where $v_{\rm{max}}$ is the maximal rotor velocity and $\tau_j$ ($\tau_1<\ldots<\tau_{14}$) are the switching times at which UAV control inputs change. Stages 1, 3 and 5 correspond to $t \in (0, \tau_4]$, $t \in (\tau_5, \tau_9]$ and $t \in (\tau_{10}, \tau_{14}]$ respectively. Whereas, stages 2 and 4 are associated with $t \in (\tau_4, \tau_5]$ and $t \in (\tau_9, \tau_{10}]$ respectively. Finally, stage 6 corresponds to $t > \tau_{14}$. The time duration for stages $s=\{1,3,5\}$ can be deduced as   
\begin{equation}
\label{eq:tau4}
\tau_{(2.5s+1.5)}-\tau_{(2.5s-2.5)}=\frac{2}{v_{\rm{max}}} \Bigg( \sqrt{\frac{\Delta \psi_{p,s} I_y}{d' \varrho}}+ \sqrt{\frac{\Delta \psi_{r,s} I_x}{d' \varrho}}\Bigg),
\end{equation}
whereas for stages $s=\{2,4\}$
\begin{equation}
\tau_{2.5s}-\tau_{(2.5s-1)}= \sqrt{2 \frac{d_{s} A_{s}}{m} },
\end{equation}
where $\tau_0=0$, $d_{s}$ is the distance traveled in stage $s$, $m$ is the UAV's mass, $\Delta\psi_{p,s}$ and $\Delta\psi_{r,s}$ are the pitch and roll changes in stage $s$, $\varrho$ is the lift coefficient, $d'$ is the distance from any rotor to the center of the UAV, and $I_x$ and $I_y$ are the moments of inertia along $x$ and $y$ directions respectively. Finally, $A_{s}$ is the total force towards the destination in stage $s$, given as $A$.

\subsection{Communication Model}

In communicating with users, BSs, or other UAVs, a certain amount of power has to be used. This power may include coding/modulation circuits, power amplifiers and frequency synthesizers when transmitting data, and low-noise amplifiers, down conversion, and demodulation/decoding circuits when receiving data \cite{Deruyck2010}. 
For simplicity's sake, in this letter, we assume that the UAV communicates with ground nodes (users and/or BSs) using power $P_{U}=\sum_{u=1}^U P_u(t)$ over time $t$, where $P_u(t)$ is the communication power to node $u$, and $U$ is the total number of nodes.

\subsection{Wireless Power Transfer Model}
DLC is a promising WPT technique for quadrotor UAVs. It is based on distributed reasoning laser. With its self-aligning feature, it can charge electric devices without specific positioning or tracking, as long as a LoS is established between the laser source and the receiver. Moreover, the small size of DLC receivers means that they can be embedded in any device, such as UAVs. Finally, a single DLC transmitter can charge several devices simultaneously \cite{Gong2013}. These advantages motivate the selection of DLC for aerial networks. Further details in \cite{Zhang2018}. 

Let $P_s(t)$ be the electrical power provided by the DLC source, while $P_0(t)$ is the maximum harvested power by the receiver on UAV in time $t$. The relationship between them is as follows \cite{Zhang2018}
\begin{equation}
\label{eq:DLC}
P_0(t)=a_1 a_2 \nu(t) P_s(t) + a_2 b_1 \nu(t) + b_2, 
\end{equation}  
where $a_1$, $a_2$, $b_1$ and $b_2$ are curve fitting parameters, $\nu(t)=e^{{-\alpha d(\textbf{w},\textbf{w}_s)}}$ is the average laser transmission efficiency, $d(\textbf{w},\textbf{w}_s)=||\textbf{w}-\textbf{w}_s||$ is the distance between the UAV and DLC source in time $t$, $\textbf{w}_s=\left[x_s, y_s, z_s\right]^t$ is the 3D location of DLC source, and $\alpha$ is the laser attenuation coefficient. The latter is expressed by $\alpha=\frac{\sigma}{\kappa} \left(\frac{\lambda}{\chi}\right)^{-\rho}$, where $\sigma$ and $\chi$ are constants, $\kappa$ is visibility factor, $\lambda$ is wavelength, and $\rho$ is the size distribution of the scattering particles. 

\section{Energy Model}
In this section, we provide the expressions of consumed and harvested energy. Consumed energy is defined by
\begin{equation}
\label{eq:con_energy}
E_{c}=E_{\rm{trav}}+E_{\rm{hov}}+E_{\rm{comm}}, 
\end{equation}
where $E_{\rm{trav}}$ is the energy to travel between locations, $E_{\rm{hov}}$ is the hovering energy, and $E_{\rm{comm}}$ is the communication energy.

The motion control energy consumed by the UAV between initial time $t_0$ and final time $t_f$ can be given by \cite{Cord1977}
\begin{equation}
\label{eq:energy}
E= \int_{t_0}^{t_f} \sum_{r=1}^4 e_r(t)i_r(t) dt,
\end{equation}
where $e_r(t)$ and $i_r(t)$ are the voltage and current across motor $r$ respectively. 
Their expressions, in steady-state conditions, are given by \cite{Morbidi2016}
\begin{equation}
    \label{eq:DCmotor}
    e_r(t)= R i_r(t) + \kappa_E v_r(t),
\end{equation}
and
\begin{equation}
    \label{eq:current}
    i_r(t)=\frac{1}{\kappa_T}\left[ T_f + \kappa_0 v_r^2(t)+D_f v_r(t)+ J \frac{\partial v_r(t)}{\partial t} \right],
\end{equation}
where $R$ is the resistance, $\kappa_E$ is the motor's voltage constant, $\kappa_T$ is the torque constant, $T_f$ is the motor friction torque, $\kappa_0$ is the drag coefficient, $D_f$ is the motor's viscous damping coefficient, and $J$ is the rotor inertia.
By combining (\ref{eq:DCmotor})-(\ref{eq:current}) into (\ref{eq:energy}), the latter can be written
\begin{equation}
\label{eq:energy2}
E = \int_{t_0}^{t_f} \sum_{r=1}^4 \Big( \sum_{i=0}^4  c_{i+1} v_r(t)^i  +  \frac{\partial v_r(t)}{\partial t} \big[ c_6 + c_7 \frac{\partial v_r(t)}{\partial t}+c_8 v_r(t) + c_9 v_r(t)^2 \big] \Big),
\end{equation}
where $c_1, \ldots, c_9$ are expressed as
\begin{eqnarray}
\label{constants}
&&c_1=\frac{R T_f^2}{\kappa_T^2}, \; c_2= \frac{T_f}{\kappa_T}\left( \kappa_E + \frac{2 R D_f}{\kappa_T} \right), \; c_3=\frac{D_f}{\kappa_T}\left( \frac{R D_f}{\kappa_T}+ \kappa_E \right) + \frac{2 R T_f \kappa_0}{\kappa_T^2}, \; c_4= \frac{\kappa_0}{T_f} c_2,\nonumber \\ \nonumber 
&&c_5=\frac{\kappa_0^2}{T_f^2} c_1, \; c_6 = \frac{2 J}{T_f}c_1, \; c_7=\frac{J^2}{T_f^2} c_1, \; c_8=\frac{J}{T_f}c_2, \; c_9=\frac{\kappa_T}{T_f}c_6. \nonumber
\end{eqnarray}
The obtained expression (\ref{eq:energy2}) will serve in determining consumed energy for the presented motion control model. Indeed, travelling energy $E_{\rm{trav}}$ can be written as
\begin{equation}
\label{eq:Efl1}
E_{\rm{trav}}=\sum_{s=1}^5 E_{s},
\end{equation}
where $E_{s}$ is the consumed energy in stage $s$ ($s = 1, \ldots, 5$). 
Since in the defined model, the motor velocities are assumed to be constant (eqs. (\ref{eq:sub1})-(\ref{eq:sub6})), the consumed energy for stages $s= \left\{1,3,5\right\}$ can be given by
\begin{equation}
\label{eq:Es1}
E_{s}=\left(\tau_{(2.5s+1.5)} - \tau_{(2.5s-2.5)}\right) \Big( 3 c_1 + (1+\sqrt{2})c_2 v_{\rm{max}}+ 2 c_3 v_{\rm{max}}^2 + (1+\frac{1}{\sqrt{2}})c_4 v_{\rm{max}}^3 + \frac{3}{2} c_5 v_{\rm{max}}^4 \Big),
\end{equation}
and for stages $s=\left\{2,4\right\}$ by
\begin{equation}
E_{s}= \left(\tau_{2.5s}-\tau_{(2.5s-1)}\right) \cdot 4 \sum_{i=1}^5 c_i v_{\rm{max}}^{i-1}.
\end{equation}
In the presence of an external force (e.g. gravity and wind), the velocity of rotors to keep the UAV aloft is given by (\ref{eq:sub6}). Hence,
the UAV's hovering energy can be written as
\begin{equation}
\label{eq:Ehv}
E_{\rm{hov}}= \Delta \cdot 4 \sum_{i=1}^5 c_i \left({\frac{|\textbf{F}_e|}{4 \varrho}}\right)^{\frac{i-1}{2}}, 
\end{equation}  
where $\Delta$ is the hovering duration. Whereas, the communication energy of the UAV is expressed by
\begin{equation}
	E_{\rm{comm}}= \int_{t_0}^{t_f} \sum_{u=1}^U P_u(t) dt. 
\end{equation}
Finally, using (\ref{eq:DLC}), harvested energy is given by
\begin{equation}
\label{eq:EEH}
E_{\rm{harv}}=\int_{t_0}^{t_f} P_0(t) dt = a_1 a_2 \nu \int_{t_0}^{t_f} \nu(t)P_s(t) dt+ a_2 b_1 \int_{t_0}^{t_f} \nu(t) dt+ b_2 \left(t_f - t_0\right). 
\end{equation}

\section{Kinetic Battery Model and Dynamics}
Since energy is either leaving or entering the battery, it is worth presenting the relation between energy and battery models. The Kinetic Battery Model (KiBaM) is adequate to model LiPo batteries \cite{Jongerden2017}. 
In KiBaM, the battery charge is divided into two wells: an available-charge well ($y_1$) and a bound-charge well ($y_2$). Given $t \in [t_0,t_f]$, and the initial battery conditions $y_1(t_0)=\omega B$ and $y_2(t_0)=(1-\omega)B$, where $B$ is the battery capacity and $\omega \in [0,1]$ is the splitting factor of well levels, the change in charge of both wells is described by the following equations \cite{Jongerden2017}
\begin{subequations}
\begin{eqnarray}
    \label{eq:well1}
    \frac{\partial y_1 (t)}{\partial t} &=& \bar{i}(t) + k_F \left( h_2(t)-h_1(t) \right)\\
    \label{eq:well2}
    \frac{\partial y_2 (t)}{\partial t}&=& -k_F \left( h_2(t)-h_1(t) \right),
\end{eqnarray}    
\end{subequations}
where $k_F$ controls the flowing rate between the wells, $h_1(t)=y_1(t)/\omega$ and $h_2(t)=y_2(t)/\left(1-\omega\right)$ are the heights of the wells, and
\begin{equation}
\bar{i}(t) = \left\{
    \begin{array}{ll}
         i_{\rm{ch}}(t) & \mbox{in the charge state}\\
        -i_{\rm{dis}}(t) & \mbox{in the discharge state},
    \end{array}
\right.
\end{equation}
where $i_{\rm{ch}}(t)$ and $i_{\rm{dis}}(t)$ are the recharge and discharge currents of the UAV's battery respectively. On one hand, we assume KiBaM constant current charging, where $i_{\rm{ch}}(t)=I_{\rm{ch}}$\footnote{Usually, charging has two phases, the first at constant maximum current until maximum voltage is reached, and the second at constant maximum voltage to keep the level of the available charge well at its maximum \cite{Jongerden2017}. Since current WPT technologies cannot recharge a flying UAV fully, only the first phase can be achieved.}. To extend the battery life, it is recommended that $I_{\rm{ch}}$ should not exceed 1C$\times B$, where 1C is a measure of the charge current, known as C-rating, and $B$ value in Ah. 
Given the nominal voltage of the LiPo battery $e_{\rm{nom}}$, harvested power respects $P_0(t) \leq P_{\rm{ch}}=I_{\rm{ch}} \times e_{\rm{nom}}$. Using (\ref{eq:DLC}), we obtain
\begin{equation}
    \label{eq:DLC2}
    P_s(t) \leq \frac{P_{\rm{ch}} - a_2 b_1 \nu - b_2}{a_1 a_2 \nu}.
\end{equation}
On the other hand, $i_{\rm{dis}}(t)=i_{\rm{cont}}(t)+i_{\rm{comm}}(t)$, where $i_{\rm{cont}}(t)=\sum_{r=1}^4 i_r(t)$ is the UAV's control current, obtained using (\ref{eq:current}), and $i_{\rm{comm}}(t)={P_U}/{e_{\rm{tr}}}$ is the communication current, where $e_{\rm{tr}}$ is the UAV transceiver's voltage.
By solving (\ref{eq:well1})-(\ref{eq:well2}) for constant $\bar{i}(t)=\Bar{I}$, we obtain the battery levels at time $t_f$ \cite{Jongerden2009}
\begin{subequations}
\begin{eqnarray}
\label{eq:y1}
y_1(t_f)&=& y_1(t_0) e^{-k' \delta } + \frac{\left( y(t_0) k' \omega + \bar{I} \right) \left( 1 - e^{-k' \delta } \right) }{k'} + \frac{\bar{I} \omega \left( k' \delta  -1 + e^{-k' \delta } \right) }{k'},\\
\label{eq:y2}
y_2(t_f)&=& y_2(t_0)e^{-k' \delta } + y(t_0)(1-\omega)\left(1-e^{-k' \delta }\right)+ \frac{\bar{I} (1-\omega) \left( k' \delta  -1+e^{-k' \delta } \right)}{k'},
\end{eqnarray}
\end{subequations}
where $k'=k_F / \left( \omega (1-\omega) \right)$, $\delta = t_f - t_0$ and $y=y_1+y_2$. 

\begin{table}[t]
\begin{center}
\caption {UAV Parameters \cite{Zhang2018,Morbidi2016,Jongerden2009}}
 \begin{tabular}{|l | l | } 
 \hline
 $\varrho$=3.8305$\cdot 10^{-6}$ N/rad/s  &  $m$=1.3 Kg   \\ [0.5ex]  \hline
 $I_x$=$I_y$=0.081 Kg.m$^2$  & $d'$=0.175 m  \\ 
 \hline
  $v_{\rm{max}}$=1047.197 rad/s  & $P_u$=0.1 W   \\
 \hline
  ($\sigma,\chi$)=(3.92, 550 nm)  & $\kappa$=3 Km  \\
 \hline
  ($a_1, b_1$)=(0.445, -0.75)  & $\rho$=0.82   \\
 \hline
($a_2, b_2$)=(0.5441, -0.231)  & $\lambda$=810 nm   \\ 
 \hline
 $\kappa_V$= 920 rpm/V  & $R$=0.2 $\Omega$  \\
 \hline
 $\kappa_E=\kappa_T$=9.5493/$\kappa_V$  & $\kappa_0=2.2518 \cdot 10^{-8}$ N.m/rad/s \\
 \hline
 $k_F=4.5 \cdot 10^{-5}$ min$^{-1}$  &  $T_f$=0.04 N.m  \\
 \hline
 $J=4.1904 \cdot 10^{-5}$ Kg.m$^2$  &  $\omega$=0.8  \\
 \hline
 $D_f$=0.0002 N.m.s/rad  & $I_{\rm{ch}}$=10 A \\
 \hline
 $e_{\rm{nom}}=3 \times 3.7= 11.1$ V & $e_{\rm{tr}}$=1 V \\
 \hline
\end{tabular}
\end{center}
\label{tab1} 
\end{table}
\vspace{-5pt}
\section{Numerical Experiment}
Based on the UAV model of Section II, we assume that a UAV flies from an initial location $\mathbf{w}_I=[100, 50, 50]$ meters to a destination $\mathbf{w}_D=[160, 80, 100]$ meters. It then hovers, communicates with ground nodes using $P_U=5$W, and recharges its battery at the same time for a duration $\Delta=20$ sec. Also, we assume that the UAV experiences gravitational and wind forces of $\textbf{F}_e = [-5, 3, -12.74]$N continuously. The DJI Phantom 2 quadrotor UAV is considered with E300 Multirotor propulsion System (2212/920KV motors) \cite{Morbidi2016}, powered by two 3-cell (3S) LiPo 11.1 V batteries with capacities $B_1=B_2$=36000 As (10Ah) \cite{Recharge}. The use of two independent batteries allows us to alternatively discharge one for motion, while recharging the other using WPT. All parameters are summarized in Table I.

\begin{figure}[t]
	\centering
	\includegraphics[width=250pt]{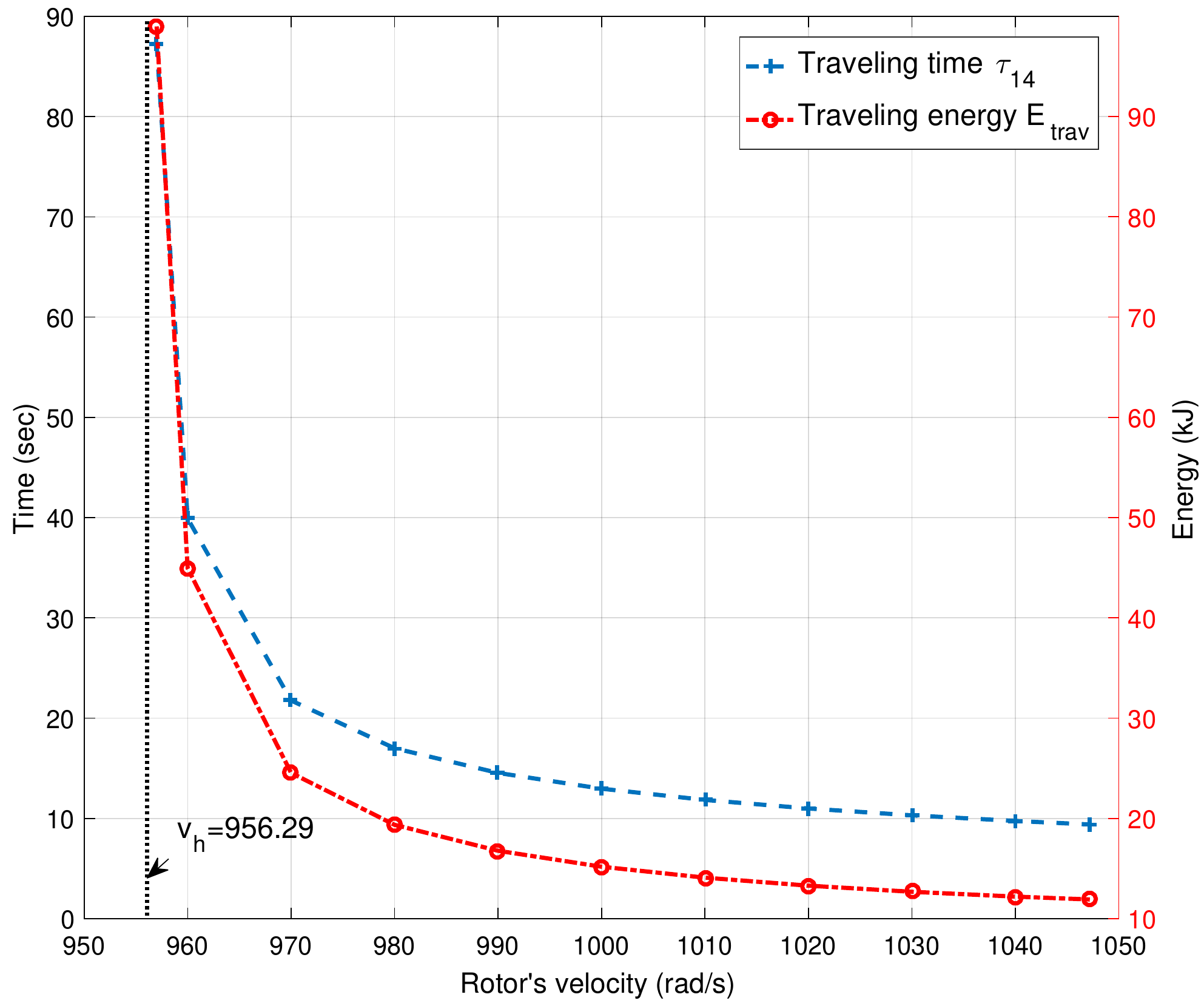}
	\caption{Traveling time and energy vs. rotor's velocity.}
	\label{Fig1c}
\end{figure}

In Fig. \ref{Fig1c}, we present traveling time $\tau_{14}$ and energy $E_{\rm{trav}}$ vs. rotor's velocity $v_r$, $r=1,\ldots,4$. Both time and energy consumption decrease with $v_r$. Indeed, a higher $v_r$ favors a faster displacement. However, for $v_r$ close to $v_h=956.29$rad/s, time and energy are very high. Indeed, $v_h$ is the hovering velocity, meaning that to move, the UAV has to provide $v_r>v_h$.

\begin{figure}[t]
	\centering
	\includegraphics[width=250pt]{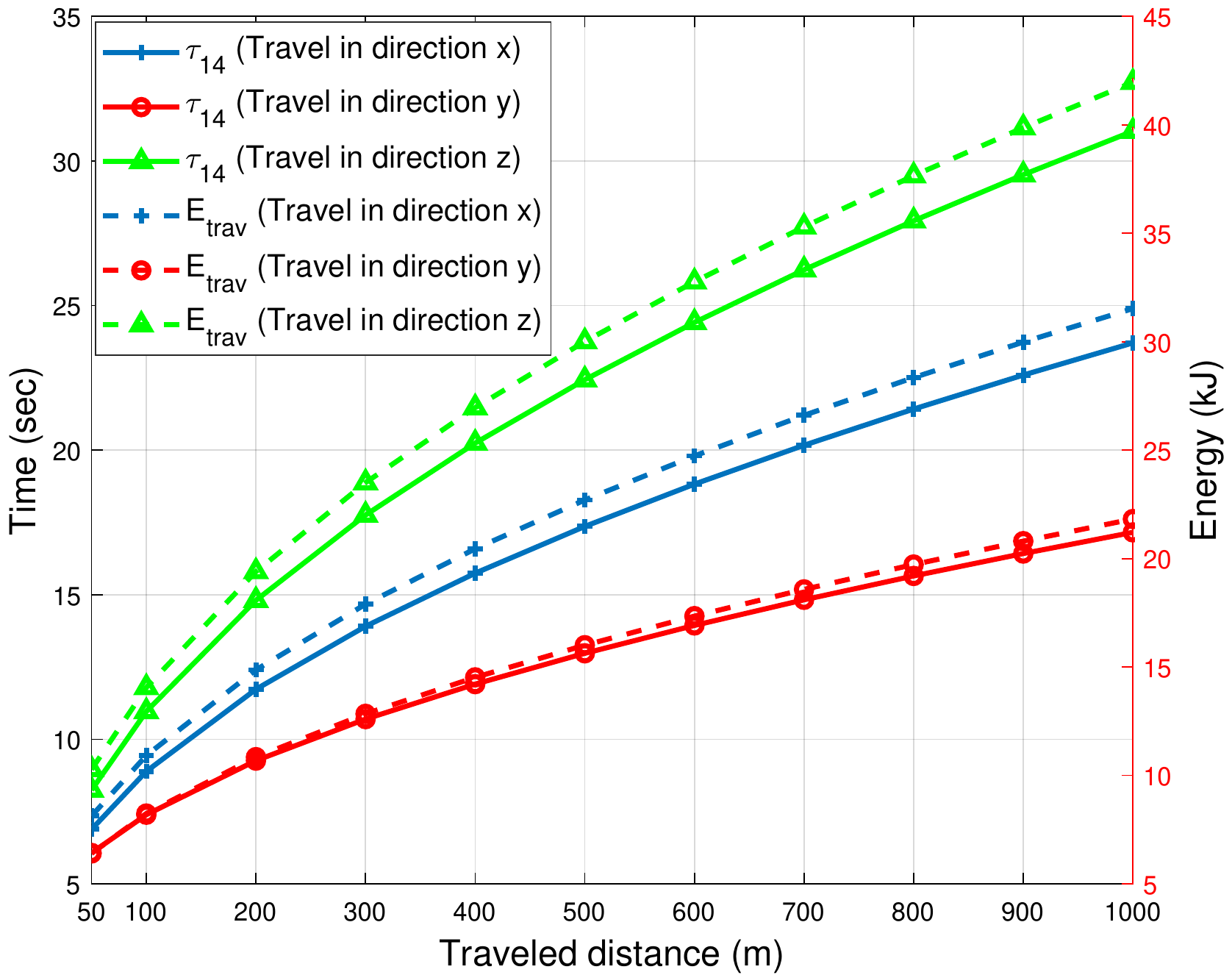}
	\caption{Traveling time and energy vs. distance.}
	\label{Fig2c}
\end{figure}
Fig. \ref{Fig2c} illustrates $\tau_{14}$ and $E_{\rm{trav}}$ vs. traveled distance in different directions in the presence of $\textbf{F}_e$. Due to the predominance of gravity, the highest amount of energy is consumed to move vertically. However, the UAV consumes less energy to move on the positive $\vec{y}$ than on $\vec{x}$. Indeed, ${F}_{e,{y}}>0$ pushes the UAV in the motion direction, which makes it use less energy. Meanwhile, ${F}_{e,{x}}<0$ is pulling the UAV back, thus forcing it to provide more forward energy.

\begin{figure}[t]
	\centering
	\includegraphics[width=250pt]{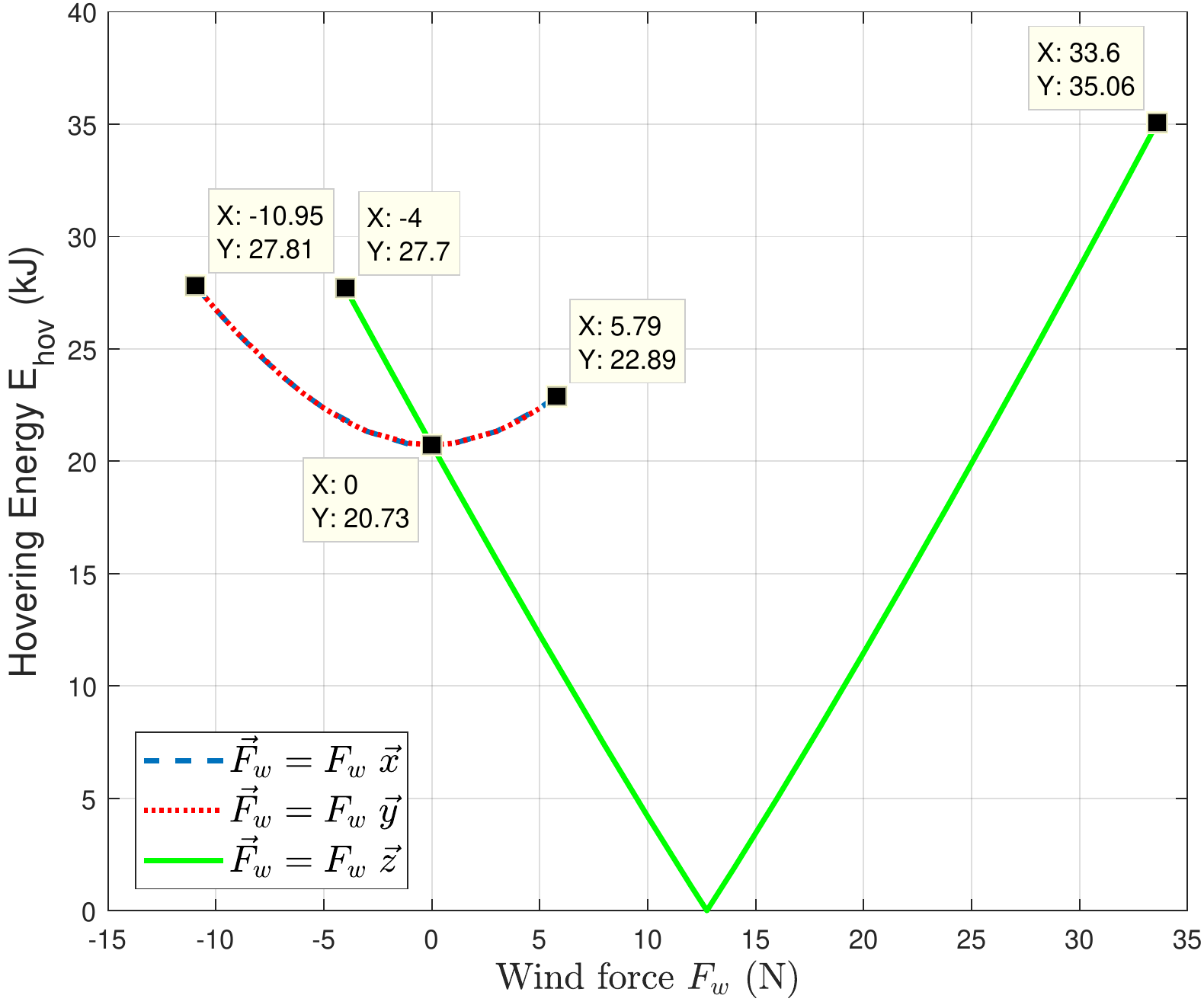}
	\caption{Hovering energy vs. wind force.}
	\label{Fig3c}
\end{figure}
Let $\vec{F}_w=\vec{F}_e - m \vec{g}$ be the wind force that hits the UAV when hovering, where $m \vec{g}$ is the gravity. The impact of $F_w$ is investigated in Fig. \ref{Fig3c}. For $\vec{F}_w=F_w \vec{x}$ (resp. $F_w \vec{y}$), $E_{\rm{hov}}$ has a parabolic shape, where the best value is for $F_w=0$N. Also, the curves are bounded by minimum and maximum $F_w$ values, corresponding to the maximum wind force that can be handled by the UAV without losing its balance.
Along $\vec{z}$, $F_w \in [0, 12.74]$N counters gravity, hence reduces $E_{\rm{hov}}$. However, for $F_w \in [-4,0]\cup[12.74,35.06]$, wind pushes the UAV to provide more force to stay aloft. Beyond these values, the UAV would lose its balance. Unlike the previous case, $E_{\rm{hov}}$ evolves linearly with $F_w \vec{z}$.

\begin{figure}[t]
	\centering
	\includegraphics[width=250pt]{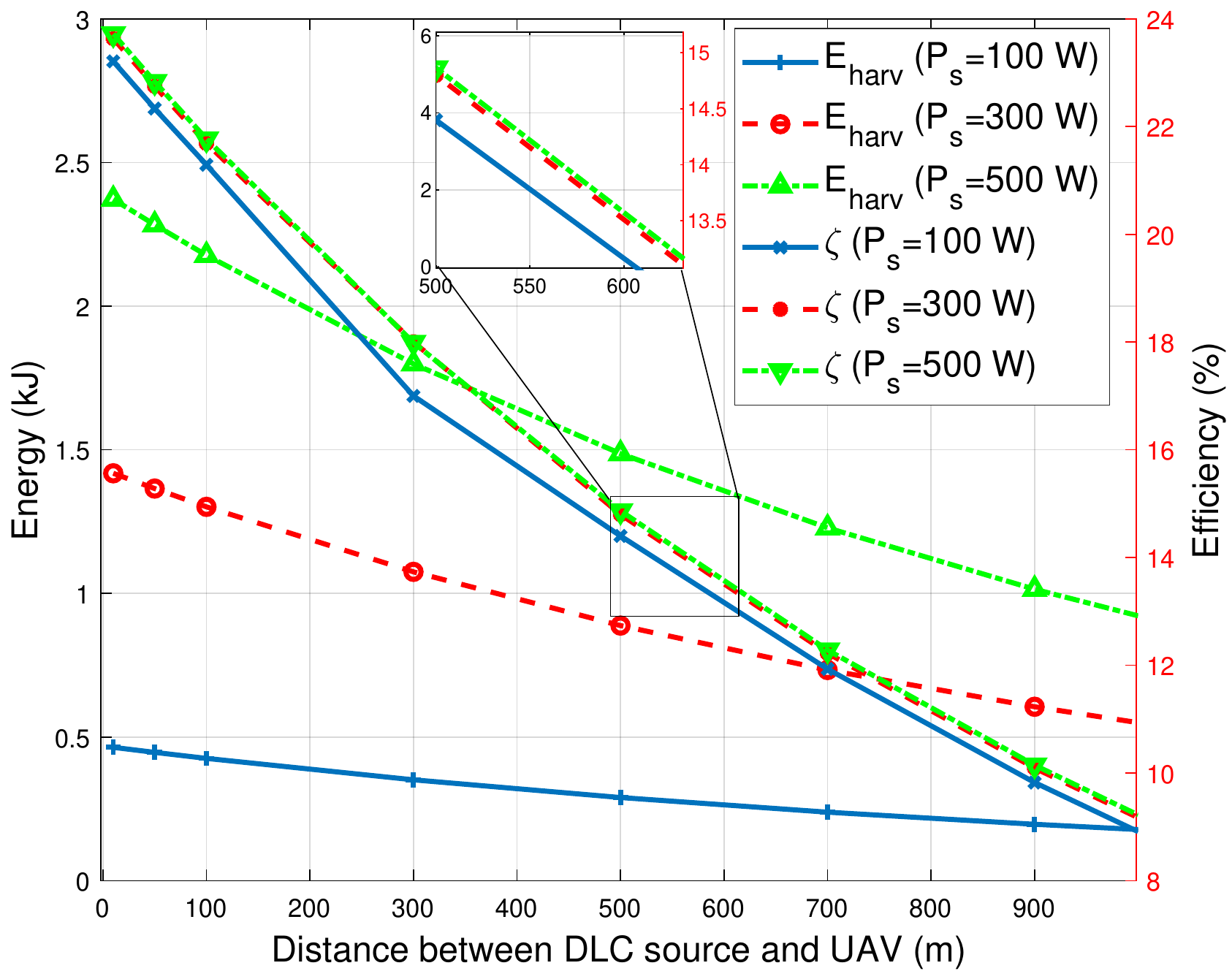}
	\caption{Harvested energy and efficiency vs. distance DLC source-UAV.}
	\label{Fig4c}
\end{figure}
Fig. \ref{Fig4c} evaluates $E_{\rm{harv}}$ and the harvesting efficiency $\zeta=\frac{P_0}{P_s}$ as functions of the distance between the DLC source and the UAV, and for different $P_s$. As the distance increases, both $E_{\rm{harv}}$ and $\zeta$ degrade due to path-loss. To improve them, $P_s$ can be increased. 

\begin{figure}[t]
	\centering
	\includegraphics[width=250pt]{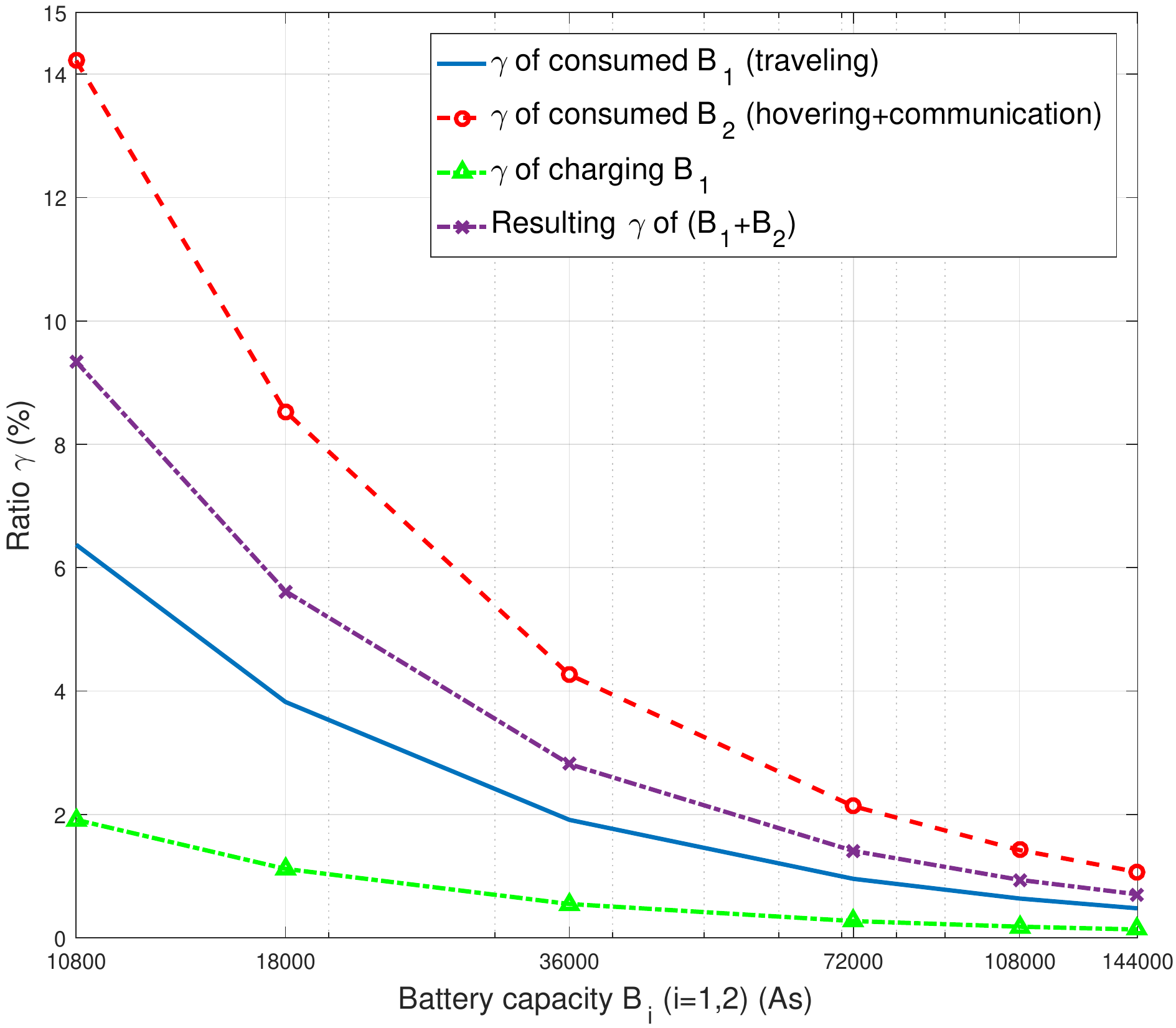}
	\caption{Consumed/Harvested battery ratio vs. battery size.}
	\label{Fig5c}
\end{figure}

In Fig. \ref{Fig5c}, we illustrate the consumed/harvested battery ratio, defined as $\gamma=1-\frac{y_1(t_f)+y_2(t_f)}{y_1(t_0)+y_2(t_0)}$ and calculated using (\ref{eq:well1})-(\ref{eq:well2}). As $B_i$ ($i=1,2$) increases, $\gamma$ decreases, since the amount of consumed/harvested energy is the same. Also, hovering+communication consumes more energy than traveling, dominated by the hovering energy. Whereas, WPT compensates for some of the lost energy. For instance, a gain of $2\%$ of $B_1=10800$As is achieved using WPT.        


\section{Conclusion}
In this letter, we proposed a simple quadrotor UAV model, where energy and battery dynamics are investigated. By leveraging the motors' and battery electrical models, we derived closed-form expressions of consumed/harvested energy and battery levels, and these were illustrated through an experiment. These results will be of great interest to researchers working on future energy-efficient aerial networks. 



%



\ifCLASSOPTIONcaptionsoff
  \newpage
\fi



%

\bibliographystyle{IEEEtran}
\bibliography{IEEEabrv,tau}

\end{document}